\documentclass[aps,twocolumn,showpacs]{revtex4-1}

\usepackage{graphicx}% Include figure files
\usepackage{dcolumn}% Align table columns on decimal point
\usepackage{bm}% bold math
\usepackage{color}% for colored comments
\usepackage{amsmath}
\usepackage{ulem}
\usepackage{flafter}

\begin{document}

\title{Systematic investigation of THz-induced excitonic Rabi splitting}

\author{M. Teich, M. Wagner, D. Stehr, H. Schneider and M. Helm}
\affiliation{Institute of Ion Beam Physics and Materials Research,
Helmholtz-Zentrum Dresden-Rossendorf,  P.O. Box 510119,  01314 Dresden, Germany}
\author{ G. Khitrova and H.M. Gibbs}\thanks{deceased}
\affiliation{College of Optical Sciences, The University of Arizona, 1630 East University Boulevard, Tucson, Arizona  85721, USA}
\author{A.C. Klettke, S. Chatterjee, M. Kira  and S.W. Koch}
\affiliation{Faculty of Physics and Material Sciences Center, Philipps University, Renthof 5, 35032 Marburg, Germany}
\date{\today}

\begin{abstract}
Weak near-infrared and strong terahertz excitation are applied to study excitonic Rabi splitting in (GaIn)As/GaAs quantum wells. 
Pronounced anticrossing behavior of the split peaks is observed for different terahertz intensities and detunings relative to the intraexcitonic heavy-hole $1s$-$2p$-transition.
At intermediate to high electric fields the splitting becomes highly asymmetric and exhibits significant broadening. 
A fully microscopic theory is needed to explain the experimental results. 
Comparisons with a two-level model reveal the increasing importance of higher excitonic states at elevated excitation levels. 
\end{abstract}
\maketitle

\section{Introduction}
The resonant optical polarization in semiconductor quantum wells (QWs) is often referred to as {\it coherent excitons} which can be classified using the well-known hydrogen quantum numbers. 
With the help of additional terahertz (THz) excitation, one can couple the different exciton states, in particular, one can induce transitions between the  $1s$ and $2p$ states \cite{Kaindl2003, Galbraith2005, Ulbricht2011, Jorger2005, Huber2006, Steiner:07, Kira2006, Kira2004}. 
In this connection, strong resonant quasi-cw THz excitation of the $1s$-$2p$ transition leads to Rabi splitting. 
Pronounced anticrossing behavior was observed by Wagner {\it{et al.}}~\cite{Wagner2010} when detuning the THz frequency from the $1s$-$2p$ transition. 
These effects were explained within the rotating wave approximation (RWA)  which already predicts  dressed states.

A number of theoretical investigations have previously been reported on the effect of a strong THz field on e.g. the excitonic absorption  \cite{Johnson:99}.
The Autler-Townes splitting and side-band generation for excitonic resonances in QWs were discussed theoretically using the RWA~\cite{Citrin1999,Liu2000}. 
However, a more extensive framework of a fully microscopic theory has already been developed \cite{Steiner:07, Kira2006, Steiner2008} to fully include non-RWA, ponderomotive, and multi-photon-ionization contributions which are needed to describe extreme nonlinear THz excitation.

\begin{figure}[b!]
 \includegraphics[width=0.4 \textwidth]{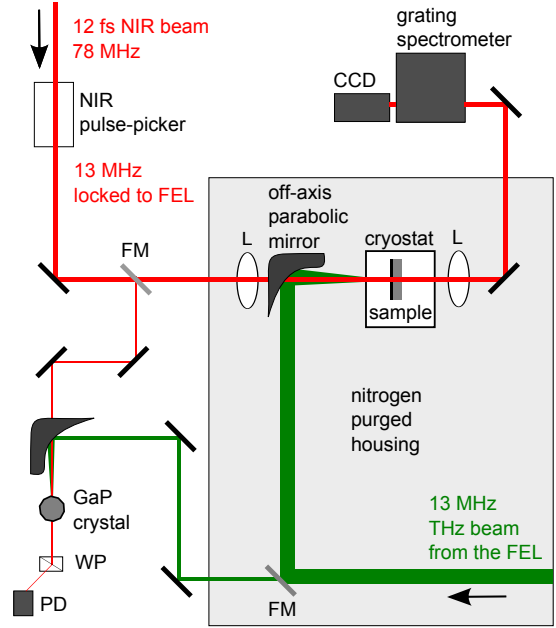}
 \caption{
 \label{Setup}
(Color online) Schematic experimental geometry. FM: flip mirror, L: lens, WP: Wollaston prism, PD: photodiode.}
 \end{figure}
 
In addition, Maslov {\it et al.}~analyzed the quantum confined Stark effect in presence of a THz field oriented in growth direction\,\cite{Maslov2000}. 
 More detailed studies concerning the combination of an ac and dc field were computed by Mi {\it{et al.}}\,\cite{Mi2004}. The interplay between the dynamic Franz-Keldysh and the ac Stark effect were also discussed by Zhang {\it et al.}~\cite{Zhang2007}. 
 Yan {\it{et al.}} investigated the interplay between ac Stark effect and dynamical localization in a semiconductor superlattice\,\cite{Yan2009}.
Rabi-splitting effects were also observed using strong THz single-cycle pulses\,\cite{Ewers2012}. 
In addition to the splitting, also pronounced resonance broadening and induced absorption effects were observed, which could be attributed to the excitation of excitonic states with high quantum numbers and THz-induced exciton ionization.

In this paper, we present a systematic investigation that is intended to fill the gap between the earlier studies and allow us to quantify the limitations of the two-level analysis that includes only the excitonic $1s$-$2p$ transition. 
For this purpose, we use the temporally long (31\,ps) THz pulses of a free-electron laser (FEL) that are spectrally sufficiently narrow to enable us to study detuning effects from the $1s$-$2p$ transition frequency. 
Furthermore, we use a sample with larger separation between heavy- and light hole energies than used in Ref.~\cite{Wagner2010}. 
We record the THz-induced changes in the near infrared (NIR) absorption spectrum for weak (1.2 kV/cm) to high THz fields up to 6 kV/cm.
Since a two-level approximation (2LA) is expected to explain only some aspects of the investigation, it is interesting to determine its validity range. As in Ref.~\cite{Ewers2012}, we use a microscopic theory~\cite{Steiner:07, Kira2006, Steiner2008} for the analysis and we compare the results to a 2LA where we restrict the treatment to the excitonic $1s$ and $2p$ states. We quantify when and how a 2LA deviates from the experimental observations, whereas the fully microscopic theory explains the experiments.

\section{Experimental Setup} 
Figure \ref{Setup} illustrates the experimental setup which is a two-beam collinear configuration for THz and NIR excitation spectroscopy. 
The setup allows the simultaneous excitation of the sample with weak broadband NIR and strong narrowband THz light.
Apart from the sample used, it is  the same setup as in Ref.~\cite{Wagner2012}.
Broadband NIR light is generated in a 12-fs Ti:Sapphire laser oscillator with a repetition rate of 78 MHz.
The repetition rate is reduced by an acousto-optical pulse picker to the 13 MHz of the free-electron laser (FEL) at the Helmholtz-Zentrum Dresden-Rossendorf.
Both beams are synchronized with a timing jitter of 1-2\,ps; the delay can be adjusted using a phase shifter.
The full width at half maximum (FWHM) of the FEL pulses is about a factor of 100 smaller than its photon energy, i.e., for a photon energy of 6.8 meV (resonant detuning) we measure 0.06\,meV FWHM yielding 31\,ps FEL pulses.
The FEL pulse duration is determined by Fourier transforming the FEL spectrum and confirmed by a cross-correlation measurement in a 300\,$\mu$m thick GaP crystal using intensity-based electro-optic sampling with a single Si photodiode \cite{CrossCorr}.
The investigated multiple quantum well sample DBR42 contains twenty 8\,nm Ga$_{0.06}$In$_{0.94}$As QWs separated by 92\,nm GaAs barriers grown on a GaAs substrate \cite{Grunwald:08}.
Unlike in the sample analyzed in Ref.~\cite{Wagner2010}, the light hole is energetically more separated from the heavy hole.

\section{Results and analysis}
\begin{figure}
\includegraphics[width=0.4 \textwidth]{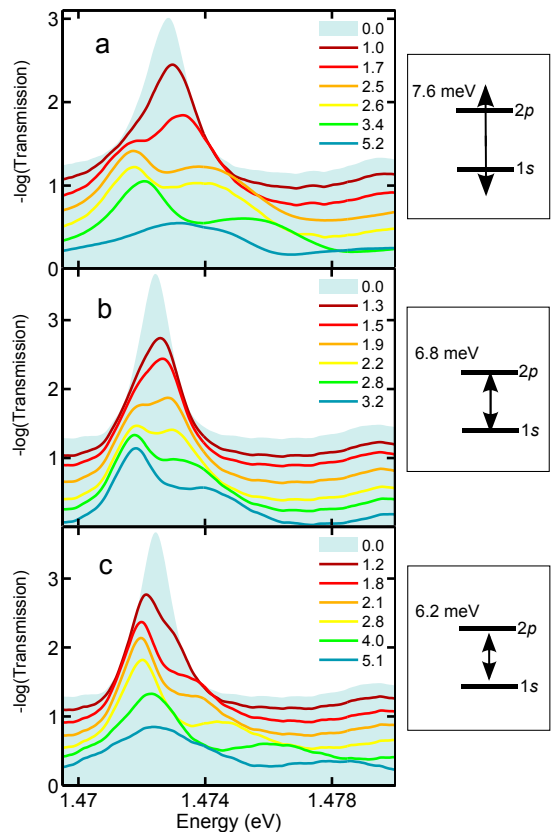}
\caption{
\label{ExpData} (Color online) Experimental absorption data (-log(Transmission)) for different THz peak field strengths (in kV/cm) and THz photon energy (a) above (7.6\,meV) (b) near (6.8\,meV) and (c) below (6.2\,meV) $1s$-$2p$ resonance.}
\end{figure} 
An overview of the measured experimental spectra plotted as -log(Transmission) is given in Fig.~\ref{ExpData} for a THz photon energy a) above, b) near and c) below the $1s$-$2p$ transition energy.
The peak field strength in kV/cm is given in the legend.
For better visibility, the spectra are vertically shifted.
In all cases, THz pumping leads to bleaching for low THz intensities, then to Rabi splitting from medium intensities on and, in Fig.~\ref{ExpData}a and b, to a reversal of the peak heights.
For high intensities, both peaks are bleached but the left peak is higher than the right peak.

\begin{figure}
 \includegraphics[width=0.45 \textwidth]{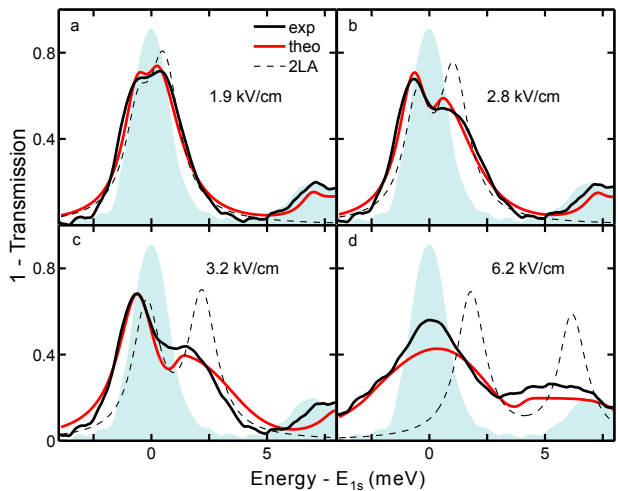}
 \caption{(Color online) 1-T without THz (shaded area) and with THz  photon energy of 6.8\,meV, near the $1s$-$2p$ resonance, as measured in experiment (black solid), fully microscopic theory (red solid) and the two-level result (dashed) for the THz field strength of (a) 1.9\,kV/cm (b) 2.8\,kV/cm (c) 3.2\,kV/cm (d) 6.2\,kV/cm.} \label{CalcNearRes}
 \end{figure}
 
We use the microscopic theory\,\cite{Steiner:07,Kira2006,Steiner2008,Ewers2012} (semiconductor Bloch equations  including THz interaction)  to treat the light-matter interaction self-consistently. 
For the THz interaction, we do not employ the RWA, but fully include the counter-rotating terms.
Besides inserting the material parameters appropriate for the sample used in the present study, we additionally evaluate the equations in the limit of the two-level approximation (2LA) where we only include the excitonic $1s$ and $2p$ states but keep all other parameters the same as in the full analysis. 
 This allows us to quantify how and when the 2LA analysis deviates from experiments.

\begin{figure}[t!]
  \includegraphics[width=0.45\textwidth]{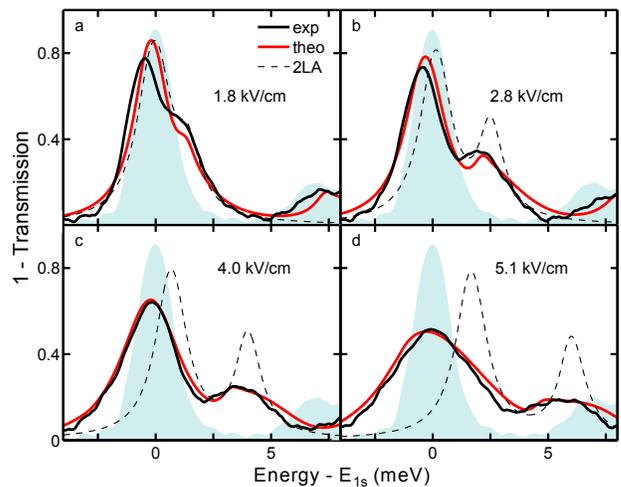}
  \caption{ (Color online) 1-T without THz (shaded area) and with THz photon energy of 6.2\,meV , below the $1s$-$2p$ resonance, as measured in experiment (black solid), from the fully microscopic theory (red solid) and
 the two-level result (dashed) for the THz field strength (a) 1.8\,kV/cm (b) 2.8\,kV/cm (c) 4.0\,kV/cm (d) 5.1\,kV/cm. }\label{CalcOffRes}
\end{figure}

Our detailed theory-experiment comparisons show that the fully theory explains all experimentally observed features. As representative examples, we show in Fig.~\ref{CalcNearRes} spectra for selected THz field strengths where the THz photon energy is close to the  $1s$-$2p$ resonance and in Fig.~\ref{CalcOffRes} for the case where the THz photon energy is tuned below the resonance energy.
Both figures present the direct comparison of the experimental data with THz (black solid) and the microscopic result (red solid). 
The experimental data without THz are shown as shaded areas, the results from the 2LA are dashed.
% ======================================
%     exp vs theo nahe der Resonanz
% ======================================

The fully microscopic theory nicely reproduces the experimental results, including the peak heights, position and THz-induced broadening.
In contrast, the 2LA works only for the lowest THz field (Fig.~\ref{CalcNearRes}a and \ref{CalcOffRes}a).
For higher THz field (2.8\,kV/cm and higher), the 2LA gets worse about the peak heights and broadening.
 For example, it overestimates the high-energy Rabi peak height by 55\,\% for 3.2\,kV/cm THz field strength for 6.8\,meV THz photon energy.
However, it roughly describes the peak splitting correctly  and the discrepancy of 2LA and experimental peak positions increases slightly with detuning.

% =====================================
%      exp vs theo off-resonant
% =====================================

\begin{figure}[b!]
 \includegraphics[width=0.55\textwidth]{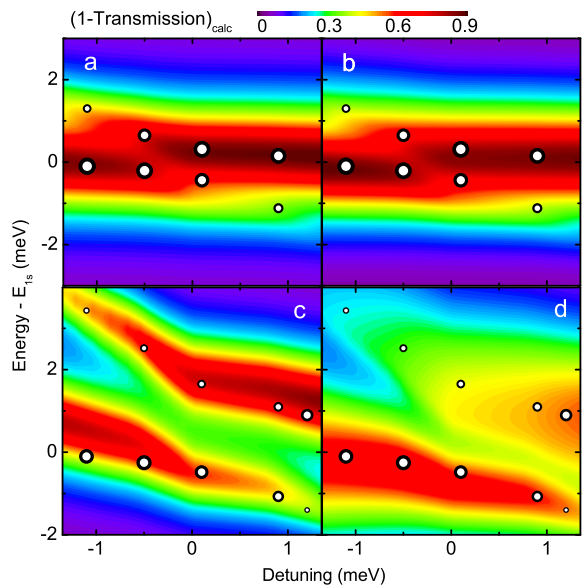}%{anticrossing}
 \caption{(Color online) Anticrossing behavior  of 1-T extracted from the microscopic calculation. 
 The experimental peak positions (heights) are depicted by the position (diameter) of the open circles. 
 (a)/(c): 2LA (b)/(d) microscopic theory results for a THz field of 1.2 (top) and 3.2\,kV/cm (bottom).}
\label{anticrossing}
\end{figure}

% -------------------------------
%   fig: occupation dynamics
% -------------------------------
\begin{figure}
 \includegraphics[width=0.35\textwidth]{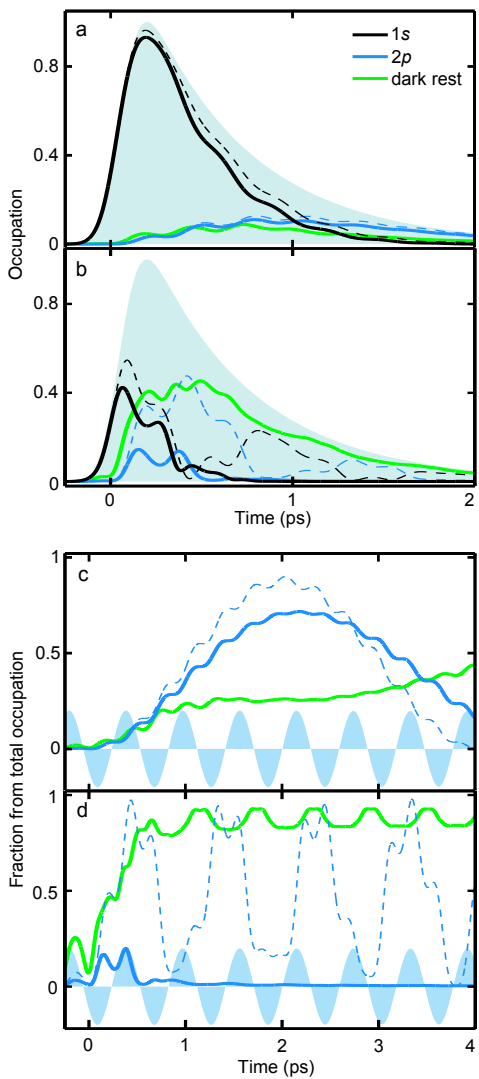}
 \caption{(Color online) \textbf{Top:} Occupations for THz photon energy near the $1s$-$2p$ resonance. The THz field strengths are (a) 1.9\,kV/cm and (b) 6.2\,kV/cm. Shaded area: Sum of all occupations without THz field. Solid/dashed lines: microscopic / 2LA result with THz field. \textbf{Bottom:}  $2p$ and the dark rest occupations as fraction of all occupations. 
Solid (dashed) lines depicts microscopic (2LA) results. Shaded area: THz field oscillations (arb. u.). Note the difference in the time scales for the upper and lower panels.} \label{PopBehav}
\end{figure}
% ---------------------------

To gain an overview of the dependence of THz field strength and detuning on the Rabi peaks, we present in Fig.~\ref{anticrossing}b and d the results for (1-Transmission) as predicted by the fully microscopic theory for low (1.2\,kV/cm) and medium THz field strength (3.2\,kV/cm), respectively. 
The experimental Rabi peaks are depicted as circles inside the contour plot, where the diameter refers to the peak height.
We observe that the Rabi splitting increases with the strength of the THz field as expected\,\cite{Kira2012}. The spectra for the 2LA are presented in Figs.~\ref{anticrossing}a and c.
For low THz field (1.2\,kV/cm, Figs.~\ref{anticrossing}a and b), the 2LA gives similar results as the microscopic theory.
However, already for a medium THz field, the microscopic theory predicts asymmetric bleaching and broadening of the Rabi peaks in agreement with the experimental results (Fig.~\ref{anticrossing}d), while the 2LA  yields symmetric splitting and bleaching (Fig.~\ref{anticrossing}c). 
Hence, the observation of anticrossing is not enough to prove the validity of a 2LA.
This shows the importance of higher excitonic states for the shapes of the $1s$-exciton peak for medium THz fields leading to a complete disagreement of 2LA and experiment for high THz fields as already seen in Figs.~\ref{CalcNearRes}d and \ref{CalcOffRes}d.

The microscopic calculations allow us to follow the redistribution of existing THz polarization also in time domain. 
Representatively, we analyze the temporal data for which the spectra were already shown in Figs.~\ref{CalcNearRes}a and d, for THz excitation near the $1s$-$2p$ resonance.
In Figs.~\ref{PopBehav}a and b,
the $1s$ occupation $|p_{1s}|^2$ without THz is plotted as shaded area as a reference.
The results for the occupations $|p_{\lambda}|^2$ from the microscopic theory with THz influence are shown in solid lines ($1s$: black; $2p$: blue; other dark states: green).
For comparison, we show the 2LA results as dashed lines.
For the low THz intensity (1.9\,kV/cm, see Fig.~\ref{PopBehav}a), the microscopic theory and the 2LA basically agree. 
However, the microscopic calculation reveals that we already have an occupation of higher lying exciton states.
For high THz field (6.2\,kV/cm, see Fig.~\ref{PopBehav}b), we see only one Rabi flop in the microscopic calculation while the 2LA incorrectly predicts multiple Rabi flopping.

To better monitor the THz influence, we replot the data for $2p$ and the dark states, but scaled as fraction of all temporarily existing occupations.
The results are shown in Figs.~\ref{PopBehav}c  and d for THz field strengths of c) 1.9\,kV/cm  and d) 6.2\,kV/cm.
For illustration, the oscillating THz field is shown by the shaded area.
For the low THz field strength (Fig.~\ref{PopBehav}c), the microscopic theory and the 2LA both describe $1s$-$2p$ Rabi flopping.
The Rabi period decreases with higher THz fields (intermediate fields not shown here) until for the highest THz field, all occupation is immediately transfered into higher dark states (see   green line in Fig.~\ref{PopBehav}d) while the 2LA incorrectly predicts continued Rabi flopping. Despite this limitations, however, the 2LA predictions of the Rabi splitting (determined by the Rabi period) are still quite correct.

\section{Conclusions}
 We have measured the THz-induced changes in the optical transmission for a THz frequency near the excitonic $1s$-$2p$ transition frequency. 
We have observed bleaching and shifting of the $1s$ exciton peak and pronounced Rabi splitting. 
In addition, the Rabi peaks undergo an anticrossing behavior for a detuning of the THz frequency away from the exciton resonance.
The fully microscopic theory excellently reproduces all experimental features and follows the transition from Rabi flopping at low THz intensities to exciton ionization at high intensities.
We also have quantified how a restriction to only two levels (2LA) deviates from experiments and the fully microscopic theory. 
The 2LA works reasonably well for low THz fields where the Rabi splitting is weak. 
For elevated THz fields, only the Rabi-peak positions are rather close to experimental values while the peak heights are not reproduced. This systematic analysis closes the gap between earlier studies showing either dressed state behaviour for low THz fields or THz ionization for strong and spectrally broad THz fields.
\section*{Acknowledgments} We thank P. Michel, W. Seidel and the FELBE team for their dedicated support. The Marburg work is supported by Deutsche Forschungsgemeinschaft. The Tucson group acknowledges support of NSF ERC CIAN and NSF AMOP.

\end{document}